\documentclass[aps,preprint,prb,showpacs]{revtex4}%
\usepackage{amsmath}
\usepackage{graphicx}%
\usepackage{amsfonts}%
\usepackage{amssymb}

\begin{document}

\preprint{ }
\title{Production of UCN by Downscattering in superfluid He$^{4}$\thanks{R.G. and
E.K. would like to thank the National Laboratory for High Energy Physics,
Tsukuba, Japan (KEK) and the Japanese Society for the Promotion \ of Science
for support.}}
\thanks{R.G. and E.K. would like to thank the National Laboratory for High Energy
Physics, Tsukuba, Japan (KEK) and the Japanese Society for the Promotion \
of Science for support.}
\author{E.Korobkina}
\author{R. Golub}
\email{golub@hmi.de}
\affiliation{Hahn Meitner Institut\\\ 14109 Berln, Germany}
\author{B.W. Wehring}
\affiliation{Nuclear Engineering Department\\North Carolina State University\\Raleigh, NC }
\author{A.R Young}
\affiliation{Physics Department\\North Carolina State University\\Raleigh, NC }

\begin{abstract}
Ultra-cold neutrons (UCN) are neutrons with energies so low they can be stored
in material bottles and magnetic traps. They have been used to provide the
currently most accurate experiments on the neutron life time and electric
dipole moment. UCN can be produced in superfluid Helium at significantly
higher densities than by other methods. The predominant production process is
usually by one phonon emission which can only occur at a single incident
neutron energy because of momentum and energy conservation. However UCN can
also be produced by multiphonon processes. It is the purpose of this work to
examine this multiphonon production of UCN. We look at several different
incident neutron spectra, including cases where the multiphonon production is
significant, and see how the relative importance of multiphonon production is
influenced by the incident spectrum.

\end{abstract}
\volumeyear{year}
\volumenumber{number}
\issuenumber{number}
\eid{identifier}
\date{\today}
\received{date}
\revised{date}
\accepted{date}
\published{date}
\startpage{101}
\endpage{102}
\maketitle
\tableofcontents

\section*{Introduction\bigskip}

The production of Ultra-Cold Neutrons (UCN) \cite{ucnbook} in superfluid
Helium and their subsequent storage allowing the build up of densities much
greater than would be achievable at thermal equilibrium with the Helium
(superthermal UCN source)\cite{bg-p} is the foundation for several current and
proposed experiments with UCN \cite{nature, physrep, lanlprop, snsprop}. Although calculations of the production rate have been
presented previously \cite{bg-p, bgskl, abe} the early
calculations concentrated on the production due to the resonant (8.9\AA)
phonons. Measurements \cite{yoshikiprl} have shown a rather narrow peak in the
production cross section as expected. In the course of the first calculation
\cite{bg-p} estimates had been made of the contribution due to higher energy
incident neutrons downscattering on the multiphonon excitation tail in
$S\left(  q,\omega\right)  ,$ and these neutrons were found to contribute
about the same amount of UCN production as the resonant neutrons for a
reasonable assumed spectrum. However as the initial work was part of a
proposal for an experiment it was felt that it would be more conservative not
to include this in the initial estimates of UCN density i.e. to keep this in
hand as a factor of safety for the predicted intensity for the new source.
However the fact that multiphonon processes would increase the UCN production
rate was mentioned.

In recent years the comparison between theory and experiment has improved
significantly and there are several proposals to operate a superthermal Helium
UCN source with a monochromatic beam of 8.9 \AA\ neutrons, so the question of
the production rate due to multiphonon processes has some practical
importance. In addition, due to a vigorous experimental program \cite{gibbs},
much more detailed information concerning $S\left(  q,\omega\right)  $ has
become available. For these reasons it has become necessary to re-examine the situation.
W. Schott, in particular, has emphasized the importance of the UCN production due to higher energy neutrons.

This work is concerned with calculations of the UCN production rate when a
volume of superfluid He$^{4}$ at temperatures of $\lesssim0.5K$ is exposed to
a neutron flux. The steady state UCN density in the source is given by%
\begin{equation}
\rho_{UCN}=P\tau
\end{equation}
where $P$ is the production rate, $UCN/cm^{2}/\sec$ and $\tau$ is the UCN
lifetime against all losses in the Helium filled storage chamber. Lifetimes
$\gtrsim500$ seconds should be achievable. The storage time in superfluid
Helium has been measured in \cite{nature, bgexp, jewell}.

\section{\textbf{Production of UCN }}

\subsection{Introduction}

A neutron at rest can absorb energy $\hbar\omega$ and momentum $\hbar q$ with
\begin{equation}
\omega=\hbar q^{2}/2m\equiv\frac{\alpha}{2}q^{2} \label{one}%
\end{equation}
with \ ($\alpha=4.14$ mev/\AA$^{-2}$) and contrarily a neutron with this
energy and momentum can come to rest after transferring its energy and
momentum to the superfluid. For single phonon interactions, which are usually
dominant, the superfluid can only exchange quantities of energy and momentum
that are related by the dispersion curve%
\begin{equation}
\omega=\omega\left(  q\right)  =cq \label{two}%
\end{equation}
where the last relation is an approximation to simplify the discussion. Then
neutrons can only come to rest by emission of a single phonon if they have the
resonant energy $E^{\ast}$ given by the intersection of equations (\ref{one})
and (\ref{two})%
\begin{align*}
\omega\left(  q\right)   &  =cq=\hbar q^{2}/2m\\
q^{\ast}  &  =2mc/\hbar
\end{align*}
In addition to one phonon interactions there are multiple phonon interactions
occurring at higher input energies that can also produce UCN. These are the
main subject of this work.

\bigskip  UCN production in Helium can be calculated as follows:
The differential cross section for neutron scattering is given by the
Fourier transform of the van Hove correlation function $S(q,\omega)$
\cite{mandL}:

\begin{equation}
\frac{d\sigma}{d\omega}=b^{2}\frac{k_{2}}{k_{1}}S(q,\omega)d\Omega
\end{equation}
and $S(q,\omega)$ has been measured in great detail \cite{gibbs}.%

\begin{equation}
d\Omega   =2\pi\sin\theta d\theta=2\pi\frac{qdq}{k_{1}k_{2}}
\end{equation}%
Then
\begin{equation}
\frac{d\sigma}{d\omega}=2\pi b^{2}\frac{k_{2}}{k_{1}}S(q,\omega)\frac
{qdq}{k_{1}k_{2}}=2\pi b^{2}S(q,\omega)\frac{qdq}{k_{1}^{2}}%
\end{equation}

Since the limits on $q$ are:

\ \ \ \ $k_{1}-k_{2}<q<k_{1}+k_{2}$

\ $k_{2}=k_{u}\ll k_{1}$ \ \ \ \ \ \ \ \ \ \ \ \ \ \ $q\sim k_{1}$ \ \ ($\hbar
k_{u}$ is the UCN momentum)

we have \ \ \ $dq=2k_{u}$. Then%

\begin{equation}
\frac{d\sigma}{d\omega} =4\pi b^{2}\frac{k_{u}}{k_{1}}S(k_{1},\omega
=\frac{\alpha k_{1}^{2}}{2})
\end{equation}
where we assumed $S(q,\omega)$ is constant over the narrow range $dq$.

$\omega=\frac{\hbar\left(  k_{1}^{2}-k_{2}^{2}\right)  }{2m}=\frac{\alpha}%
{2}\left(  k_{1}^{2}-k_{2}^{2}\right)  \approx\frac{\alpha}{2}k_{1}^{2}$

\ \ \ \ \ \ \ \ \ The UCN production rate is given by%

\begin{equation}
P(E_{u})\bigskip dE_{u}=\left[  \int\frac{d\Phi\left(  E_{1}\right)  }%
{dE}N_{He}\frac{d\sigma}{d\omega}%
\left(  E_{1}\rightarrow E_{u}\right)  dE_{1}\right]  dE_{u}%
\end{equation}

\bigskip 
then%

\begin{align}
\int_{0}^{E_{c}}P(E_{u})\bigskip dE_{u}  &  =N_{He}4\pi b^{2}\alpha^{2}\left[
\int\frac{d\Phi\left(  k_{1}\right)  }{dE}S(k_{1},\omega=\frac{\alpha
k_{1}^{2}}{2})dk_{1}\right]  \int_{0}^{k_{c}}k_{u}^{2}dk_{u}\nonumber\\
&  =N_{He}4\pi b^{2}\alpha^{2}\left[  \int\frac{d\Phi\left(  k_{1}\right)
}{dE}S(k_{1},\omega=\frac{\alpha k_{1}^{2}}{2})dk_{1}\right]  \frac{k_{c}^{3}%
}{3}\quad UCN/cm^{3}/\sec\label{111}%
\end{align}
where $E_{c}$, $k_{c}$ are the critical UCN energy and wave vector of the
walls of the storage chamber and $\frac{d\Phi\left(  k_{1}\right)  }{dE}%
=\frac{d\Phi\left(  E_{1}=\frac{\alpha k_{1}^{2}}{2}\right)  }{dE}$ is the 
energy spectrum of the incoming flux.

Note $S(k_{1},\omega)$ is in $\left(  mev\right)  ^{-1}.$%

\subsection{Single phonon production of UCN}

The one phonon production rate \cite{ucnbook, bg-p, jewell} is found by evaluating the integral in equ.(\ref{111})
over the one phonon peak. (Note that $k^{\ast}=.7$ \AA$^{-1}$) .

\begin{equation}
P_{1p} =N4\pi b^{2}S^{\ast}\alpha\beta\frac{k_{c}^{3}}{3k^{\ast}}%
\frac{d\Phi\left(  E_{1}^{\ast}\right)  }{dE}\quad UCN/cm^{3}/\sec\label{100}%
\end{equation}
where $S^{\ast}=S_{1}\left(  k^{\ast}\right)  =.1$ \cite{gibbs, c and
w}\ and $\beta=\frac{v_{n}^{\ast}}{v_{n}^{\ast}-c_{g}^{\ast}}=1.45$
\cite{ucnbook, jewell, pend priv}\ with $v_{n}^{\ast}$ the neutron velocity at the
critical energy for UCN production, $E^{\ast}$, and $c_{g}^{\ast}$ the phonon
group velocity at the critical energy.
We have replaced $\int_{P}S\left(
q,\omega=\ \frac{\alpha}{2}q^{2}\right)  d\omega$ where the subscript means
the integration is carried out across the phonon peak with $\beta S_{1}\left(
k^{\ast}\right)  $ where
\begin{equation}
S_{1}\left(  k^{\ast}\right)  =\int_{P}S\left(  k^{\ast},\omega\right)
d\omega\label{S1}%
\end{equation}
is the contribution to the structure factor from the one phonon scattering \cite{gibbs, c and w} and $\beta$ represents the increase in the integral due to the
fact that the two curves (\ref{one}) and (\ref{two}) cross at an angle, so the
path of integration is longer than in the calculation of (\ref{S1}), as was
first pointed out by Pendlebury\cite{pend priv}. Numerical integration of the
measured $S\left(  q,\omega=\ \frac{\alpha}{2}q^{2}\right)  $ \cite{gibbs}
across the phonon peak, (fig. 1) confirms this procedure. (Note that the
factor of 2 in equation 9 of [\onlinecite{bg-p}], which holds in the case of a linear
dispersion relation should be replaced by $\beta=1.45$ for the real,
non-linear dispersion relation)

Thus\ \ \ (see equation \ref{100})\ \ \ \ \ \ \ \ \ \ \ \ \ \ \ \ \ \ \ \ \ \ \ \ \ \ %

\begin{equation}
P_{UCN}=9.\,\allowbreak44\times10^{-9}\frac{d\Phi\left(  E_{1}^{\ast
}\right)  }{dE_{1}^{\ast}}\quad UCN/cm^{3}/\sec\label{101}%
\end{equation}
where $\frac{d\Phi\left(  E_{1}^{\ast}\right)  }{dE_{1}^{\ast}}$ is in units
of $n/cm^{2}/\sec/mev$.

\subsection{Numerical calculations of the multiphonon down scattering cross section}

Fig. \ref{fig1} shows $S(q,\omega=\frac{\alpha q^{2}}{2})$ as a function of \ $q$
obtained from [\onlinecite{gibbs}].  We
have extrapolated the data above $1.2\quad$\AA$^{-1}$ in fig. \ref{fig1}, taking into
account the known value of $S\left(  q\right)  $ for the one phonon and
multiphonon contributions \cite{c and w, gibbs}. Measurements
\cite{fak} show that $S(k_{1},\omega=\frac{\alpha k_{1}^{2}}{2})$ is
essentially zero for $k_{1}\gtrsim2\quad$\AA$^{-1}$.

As an illustration we apply the above technique to various measured and
calculated slow neutron spectra that are under discussion as locations for
various superthermal UCN source projects. We will take $k_{c}=\allowbreak
1.\,\allowbreak1\times10^{-2}\quad$\AA$^{-1}$ corresponding to the critical
energy for Beryllium.

\subsubsection{North Carolina State proposed UCN source}%

\begin{figure}
[ptb]
\begin{center}
\includegraphics[bb= 100 500 460 870,
height=3.7066in,
width=5.047in
]%
{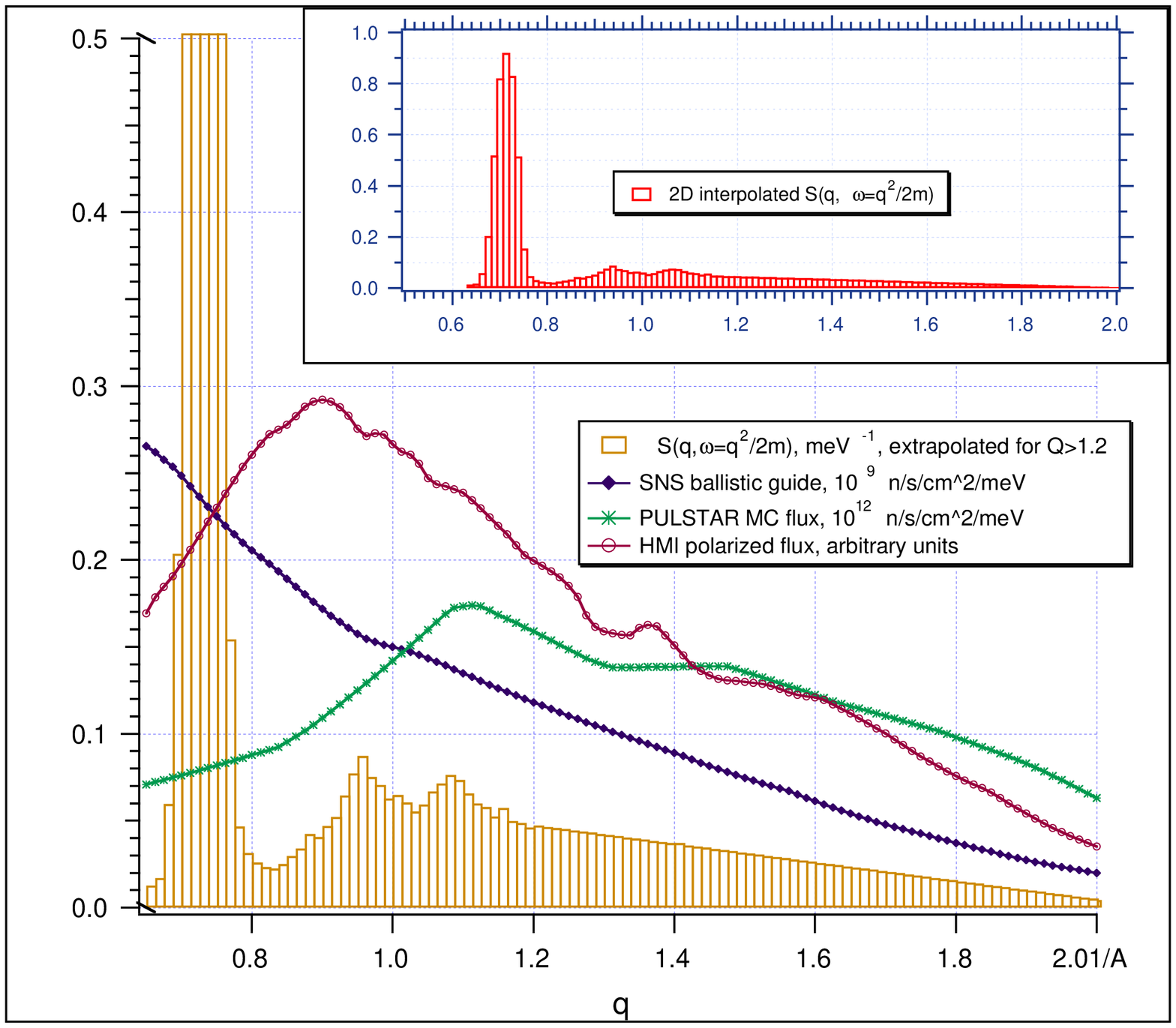}%
\caption{\label{fig1}The energy spectrum of the neutron flux from three sources compared to
the scattering law, $S(q,\omega=\frac{\alpha q^{2}}{2})\quad mev^{-1}.$ as a
function of q, \AA$^{-1}.$}
\end{center}
\end{figure}

North Carolina State University researchers have proposed to locate a UCN
source in the thermal column of the campus 1-MW PULSTAR reactor after removing
the graphite. The expected average cold neutron flux in the UCN converter was
estimated using MCNP simulations with a detailed model of the PULSTAR core and
a conceptual model of the UCN source. The core's 4 control blades and 25 fuel
assemblies, each consisting of a 5 x 5 array of fuel pins containing 4 \%
enriched UO2, were faithfully represented. The model of the UCN source
consisted of a cold neutron source surrounding a UCN converter about 10-cm
deep in a tank of D2O in the thermal column void. The cold source moderator
was 1-cm-thick cup-shaped solid methane (408 g, 22 K) contained in a
2-mm-thick-wall aluminum chamber (1164 g). The UCN converter was modeled by a
disk of liquid ortho-deuterium (4-cm thick x 14-cm diameter, 20 K) in a
5-mm-thick-wall aluminum container (484 g). Neutrons leaving a bare face of
the reactor core were channeled into the D$_{2}$0 tank by a 44-cm diameter x
75-cm long void in a beryllium assembly located between the reactor core and
the reactor tank wall. The simulated spectrum is shown in fig. \ref{fig1}, along with
the spectrum of some other sources discussed below and \ $S(q,\omega
=\frac{\alpha q^{2}}{2})$.

For this spectrum

$\left[  \int\Phi\left(  E_{1}\right)  S(k_{1},\omega=\frac{\alpha k_{1}^{2}%
}{2})dk_{1}\right]  =5.22\times10^{9}$

Substituting into equation (\ref{111}):

$\left(  N=2.18\times10^{22}\right)  $

we have%
\[
P_{mp}=N\times4\pi b^{2}\times5.22\times10^{9}\times\alpha^{2}\frac{k_{c}^{3}%
}{3}=984\quad UCN/cm^{3}/\sec
\]

For the one phonon contribution we have, taking $\left.  \frac{d\phi}%
{dE_{mev}}\right|  _{E^{\ast}}$ from fig. \ref{fig1} and using equation (\ref{101})%

\[
P_{1p}=9.\,\allowbreak44\times10^{-9}\left.  \frac{d\phi}{dE_{mev}%
}\right|  _{E^{\ast}}=9.\,\allowbreak44\times10^{-9}\times7.9\times
10^{10}=745\quad UCN/cm^{3}/\sec
\]
These figures will be reduced by a factor of 2 when the flux (calculated here
for a modest source volume as described above) is averaged over a more
reasonable 20 liter volume. The effects of construction materials in a
realistic design have not yet been taken into account.

The advantages of such a design were discussed in [\onlinecite{bgtriga}]. The main
point is that the UCN production process is independent of the direction of
the incident neutrons so that use is made of the full solid angle of the flux.
Installations at a beam position, as discussed in the next section, lose a
solid angle factor of about 10$^{4}$ [\onlinecite{bg-p}] in comparison to
installations inside a 4$\pi$ flux, although such sources outperform any other
type of UCN source that could be installed at the same beam position.

\subsubsection{Proposed UCN source at the Spallation Neutron Source (SNS)}

\begin{figure}
[ptb]
\begin{center}
\includegraphics[
height=3.9306in,
width=5.047in
]%
{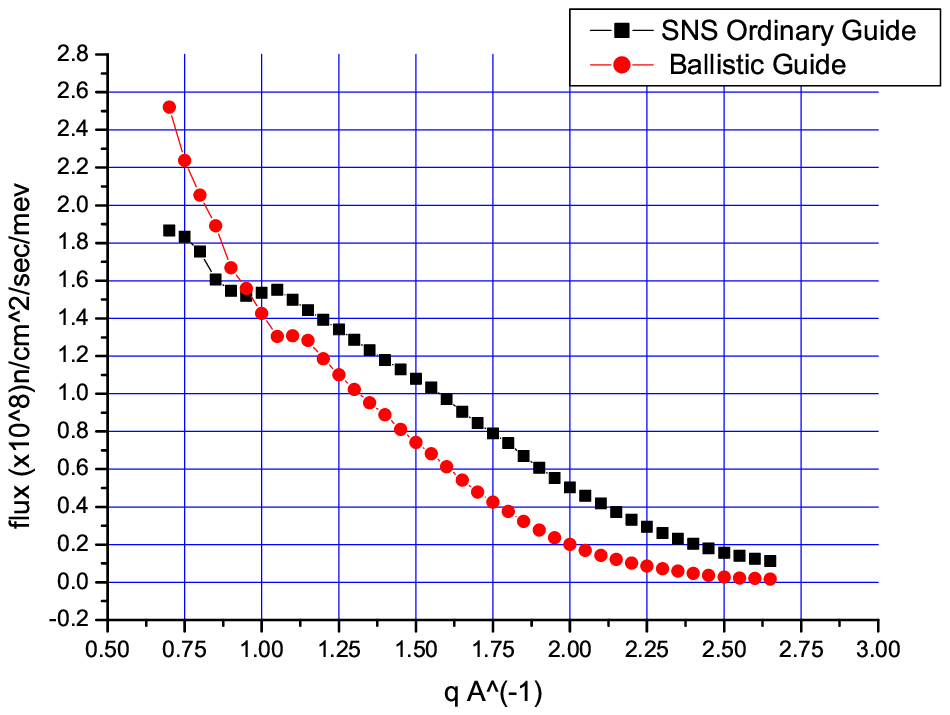}%
\caption{\label{fig2}Comparison of simulated spectra from the two types of guide proposed
for the SNS fundamental physics position as a function of q, \AA$^{-1}.$}
\end{center}
\end{figure}

It has been proposed to place a superthermal source of UCN on a monochromatic
8.9\AA\ beam at a guide tube at the SNS.\cite{snsprop} It is interesting to
ask what the loss of UCN production rate associated with the use of a
monochromatic beam would be in the case of the simulated spectra available for
the planned guide tubes. Spectra have been simulated for two types of guide:
an 'ordinary' supermirror guide and a so-called 'ballistic' guide using
tapered sections to focus and defocus the beam at the borders of a long
free-flight region. This is an interesting case for the present discussion as
the 'ballistic' guide has more flux at the critical 8.9\AA\ wavelength but
less flux at shorter wavelengths ($\lambda<6$\AA). The predicted fluxes for
the two types of guide are shown in fig. \ref{fig2}, while fig. \ref{fig1}  shows a comparison of
the SNS ballistic guide with the other sources considered in this paper.%

The result for the 'ordinary' guide is that
\[
\int\Phi\left(  E_{1}\right)  S(k_{1},\omega=\frac{\alpha k_{1}^{2}}{2}%
)dk_{1}=5.38\times10^{6}%
\]

so that
\[
P_{mp}=N\times4\pi b^{2}\times5.38\times10^{6}\times\alpha^{2}\frac{k_{c}^{3}%
}{3}=1.\,\allowbreak015\quad UCN/cm^{3}/\sec
\]

For the one phonon production rate we find using (\ref{101})%
\[
P_{1p}=9.\,\allowbreak44\times10^{-9}\left.  \frac{d\phi}{dE_{mev}%
}\right|  _{E^{\ast}}=1.\,\allowbreak8\quad UCN/cm^{3}/\sec
\]

For the 'ballistic' guide we have
\[
\int\Phi\left(  E_{1}\right)  S(k_{1},\omega=\frac{\alpha k_{1}^{2}}{2}%
)dk_{1}=5\times10^{6}\quad
\]
so that $P_{mp}=\allowbreak0.94$ and the one phonon production rate is:
$2.\,\allowbreak36\quad UCN/cm^{3}/\sec$.

Thus we see that the losses associated with rejecting the higher energy
neutrons in the beam by the monochromator are less significant than the losses
(50\%) \cite{snsprop} connected with the monochromator and associated beam bender.

\subsubsection{HMI beam spectrum}

This spectrum was measured at the NL4 neutron guide at the Hahn Meitner
Institut, Berlin by time of flight. In this context we should remind the
reader that the measurement of the spectrum from a neutron guide is quite
complex as the spectra in general are very strong functions of the flight
direction and position in the beam cross section. For the measured spectrum
(normalized to 1 at the peak of $d\phi/d\lambda$) we find:%
\[
\left[  \int\Phi\left(  E_{1}\right)  S(k_{1},\omega=\frac{\alpha k_{1}^{2}%
}{2})dk_{1}\right]  =.025
\]
yielding%
\[
P_{mp}=4.\,\allowbreak72\times10^{-9}%
\]

the one phonon contribution for this spectrum is found (equ. \ref{101}) as
\[
P_{1p}=5.\,\allowbreak48\times10^{-9}%
\]%

\begin{figure}
[ptb]
\begin{center}
\includegraphics[bb= 100 600 460 920,
height=3.0493in,
width=4.9562in
]%
{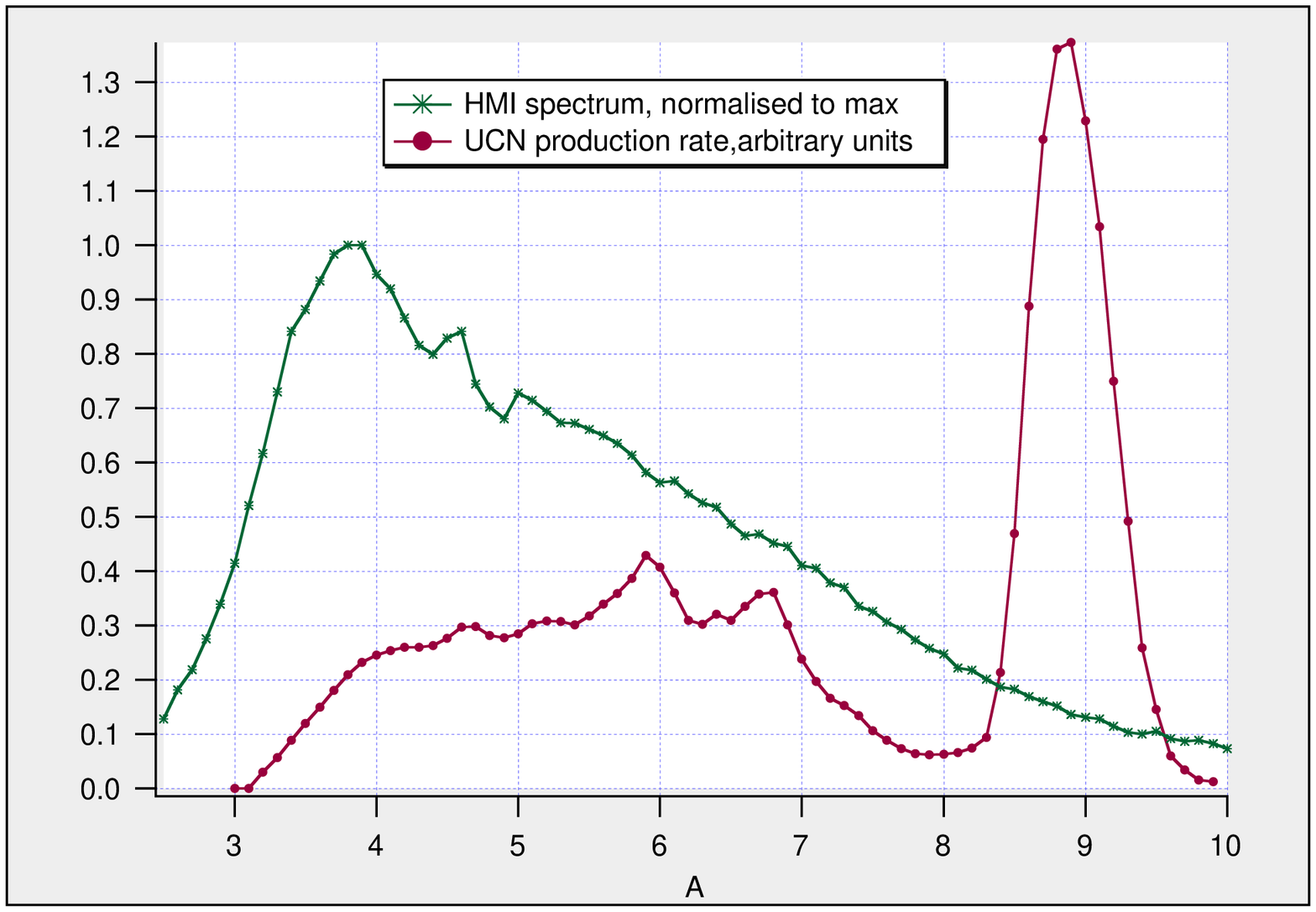}%
\caption{\label{fig3}The energy spectrum of the flux measured on the beam NL4 of the HMI
and the calculated UCN production as a function of $\lambda-$\AA.}
\end{center}
\end{figure}

Fig. \ref{fig3} shows the spectrum and calculated UCN production rate as a function of
the wavelength of the incident neutrons, $\lambda$, similar to what would be
observed in a time of flight measurement.

\section{Discussion}

The results are summarized in the table below. The column labelled 'Maxwell''
refers to a Maxwellian spectrum corrected for an ideal guide transmission
$\tau\propto\lambda^{2}$ and cut off at $3.8$\AA\ minimum wavelength.

We see that the multiphonon contribution as expected is a rather strong
function of the source spectrum varying from less than to slightly more than
the one phonon production rate for the realistic spectra considered here.

\bigskip

\textbf{Table 1 - Predicted production rates \ \ \ \ \ }%

\begin{tabular}
[c]{|l|l|l|l|l|l|}\hline
& NC State$^{\ast}$ & SNS ord$^{\ast}$ & SNS ball$^{\ast}$ & HMI a.u. &
Maxwell\\\hline
multi-ph & 490 & 1.0 & .94 & 4.7 & 1.7\\\hline
single ph & 375 & 1.8 & 2.4 & 5.5 & 1.5\\\hline
mph/1ph & 1.4 & .55 & .4 & .85 & $\allowbreak$1.\thinspace\allowbreak
13\\\hline
\end{tabular}

$^{\ast}$\textbf{UCN/cm}$^{3}$/sec

(The NC State figures have been reduced by the factor of 2 for a 20 liter
source volume as discussed above.)

Thus we see that the inclusion of the multiphonon production amounts to at
most a little more than a factor of 2 increase in UCN production. In the case
of a cold beam the multiphonon contribution is a small correction to the
single phonon production so that the use of a monochromatic beam, which offers
significant operating advantages \cite{snsprop}, would be accompanied by a
minor loss in UCN production.

In contrast, for sources where the Helium is exposed to the total thermal flux
as in the North Carolina State proposal or at a dedicated spallation source,
the multiphonon contribution can amount to slightly more than a factor of 2
increase in UCN\ production.

\begin{acknowledgments}
We are extremely grateful to Ken Andersen for sharing his data and for his
patient explanations.
\end{acknowledgments}

\end{document}